
\magnification=\magstep1
\baselineskip=20pt
\centerline{Two Classes of Gamma-Ray Bursts}
\medskip
\centerline{J. I. Katz and L. M. Canel}
\centerline{Department of Physics and McDonnell Center for the Space
Sciences}
\centerline{Washington University, St. Louis, Mo. 63130}
\bigskip
If gamma-ray bursts are at cosmological distances, as suggested by their
isotropy on the sky and the comparative deficiency of weak bursts, then they
represent radiated energies of $\sim 10^{51}$ erg, and imply the release of
an even greater energy.  Only neutron stars and black holes have binding
energies sufficient to power such extraordinarily violent and energetic
events.  General considerations of neutrino opacity imply$^1$ that the
escape of a neutron star's (or black hole's) binding energy requires a time
of about 10 sec, as shown by the observed duration of neutrino emission from
SN1987A.  The distribution of durations of gamma-ray bursts is known$^2$ to
be bimodal, with one peak between 10 and 100 sec and the other between 0.1
and 1 sec.  We hypothesize that the durations of the longer bursts may be
explained as the result of the diffusion of energy, by means of neutrinos,
from a forming neutron star or black hole, but that the brevity of the
shorter bursts requires different physics.  An alternative hypothesis
supposes that all bursts (excepting soft gamma repeaters, which we do not
discuss) represent a single class of events, whose differing durations
reflect differences in one or more parameters.  These two hypotheses may be
tested using data from the recently released 3B Catalogue$^3$.

We first consider the spectral behavior of gamma-ray bursts.  If they are a
single class of events, we may expect the spectra of long and short bursts,
measured across the entire band over which data are available (from a few
tens of KeV to many MeV), to follow a single pattern.  If they represent two
distinct classes of events, with distinct physical mechanisms and
properties, then long and short bursts may have qualitatively different
behavior.  The 3B Catalogue contains for most bursts a spectral hardness
ratio which is the ratio of the count rates in two detectors whose nominal
sensitivity bands are 100--300 KeV and 50--100 KeV; it is a measure of the
spectral slope in soft gamma-rays.  A small fraction of the bursts are also
detected by instruments sensitive to more energetic gamma-rays: COMPTEL
(nominally 1--30 MeV), OSSE (0.06--10 MeV) and EGRET (20-30,000 MeV).

The data are summarized in Table 1.  We divide the bursts into two classes,
short and long, on the basis of whether the duration parameter $T_{90}$
(the time during which the middle 90\% of the fluence arises) is less or
more than 10 sec, respectively.  Bursts with BATSE hardness ratio $> 10$
have unusually hard spectra in the soft gamma-ray range; it is these bursts
which would be expected to appear in the higher energy detectors if spectral
behavior could be simply extrapolated to higher energy.

The data refute this.  Almost all the bursts with BATSE hardness ratio
$>10$ are short, while nearly all the bursts detected by COMPTEL (and all
detected by OSSE and EGRET) are long.  This establishes that short and long
bursts have qualitatively different spectral behavior.  The probability that
the data shown in the first two lines of Table 1 (BATSE and COMPTEL) could
have been drawn from a single population of events is $< 10^{-8}$.  The OSSE
and EGRET data are consistent with the COMPTEL data.

This conclusion can be described in a different way: It is known$^2$ that
short bursts have, on average, larger BATSE hardness ratios than long
bursts.  It would therefore be expected that short bursts are more likely to
be detected at higher photon energies than long bursts.  Table 1 shows that
the opposite is true.  The higher energy photons must be produced by a
process which acts in long bursts but not in short ones, establishing that
the two classes of bursts differ in more than their durations (or parameters
correlated with duration).

The extraordinary burst 3B940217$^4$ is illustrative of this conclusion,
although as a single event it is not statistically significant on its own.
It was very long ($T_{90} = 150$ sec), but produced the most energetic
photon (18 GeV) ever observed from a gamma-ray burst, as well as many other
photons of $\sim 100$ MeV energy.

The spatial distribution of long and short bursts may also be compared.  We
find, using the 3B Catalogue, that each group is isotropically distributed
on the sky to within the limits of statistical accuracy: neither
distribution has a significant dipole or quadrupole moment, and neither shows
a significant autocorrelation at small angular scales, implying no
detectable repetitions.  These results confirm those obtained earlier$^2$
with a smaller sample of data.

The 3B Catalogue contains 251 short bursts with values for $C/C_{min}$,
where $C$ is the peak count rate and $C_{min}$ the detection threshold, and
365 long bursts with values for $C/C_{min}$; the remaining 506 bursts lack
$T_{90}$ or $C/C_{min}$.  Using these data we evaluated the parameter$^5$
$\langle V/V_{max} \rangle$ separately for short and long bursts, with the
results $0.385 \pm 0.019$ for short bursts and $0.282 \pm 0.014$ for long
bursts.  Each of these results is significantly less than 0.5, as previously
noted$^2$ for a smaller sample.  Combined with the observed isotropy, this
implies that each class of burst is at cosmological distances (or, possibly,
in an extended Galactic halo).  Our new result is that the difference in the
values of $\langle V/V_{max} \rangle$ between short and long bursts is
$0.103 \pm 0.024$.  This is significant at the $4.3\,\sigma$ level, very
unlikely to have arisen as a statistical fluctuation, and implies that long
and short bursts have different spatial distributions.

The observation of this difference in spatial distribution between short and
long bursts supports the conclusion that they are different kinds of events.
Because this is a quantitative rather than a qualitative difference, it
might also be explained if there were a single class of burst whose duration
depended on one (or more) parameters, which varied systematically with
distance.  However, this explanation would not be easy to reconcile with
the distinctly bimodal$^2$ distribution of $T_{90}$; it is more plausible to
take this result as supporting the conclusion, drawn from the spectral data,
that long and short bursts represent fundamentally different classes of event.

Despite their spectral and statistical differences, long and short bursts
qualitatively resemble each other in many ways.  Both classes have
nonthermal soft gamma-ray spectra, usually with spectral turn-overs in the
100--300 KeV range$^6$.  Members of both classes often have complex
multi-peaked time profiles, and rapidly rising but more slowly decaying
subpulses$^7$ are common.  Although we have argued that long and short
bursts must have fundamentally different mechanisms, their similarities are
obvious.  This phenomenological similarity suggests that the soft
gamma-rays of both classes are radiated by similar processes.  Only the
underlying mechanisms determining their energy release and time scales
differ.

This research has made use of data obtained through the Compton Gamma-Ray
Observatory Science Support Center Online Service, provided by the
NASA-Goddard Space Flight Center.  We thank R. Kippen and C. Kouveliotou for
discussions and NASA and NSF for support.
\vfil
\eject
\centerline{References}
\bigskip
\parindent=0pt
1. Dar, A., Kozlovsky, B. Z., Nussinov, S. \& Ramaty, R. {\it Astrophys. J.}
{\bf 388}, 164--170 (1992).

2. Kouveliotou, C. {\it et al. Astrophys. J.} {\bf 413}, L101--L104 (1993).

3. Meegan, C. A. {\it et al. Astrophys. J. Suppl.} in press.

4. Hurley, K. {\it et al. Nature} {\bf 372}, 652--654 (1994).

5. Schmidt, M., Higdon, J. C. \& Hueter, G. {\it Astrophys. J.} {\bf 329},
L85--L87 (1988).

6. Schaefer, B. E. {\it et al. Astrophys. J. Suppl.} {\bf 92}, 285--310
(1994).

7. Fishman, G. J. {\it et al. Astrophys. J. Suppl.} {\bf 92}, 229--283
(1994).
\vfil
\eject
\halign{\quad#\hfil\quad&\quad\hfil#\hfil\quad&\quad\hfil#\hfil\quad\cr
&$T_{90} < 10$ sec&$T_{90} > 10$ sec\cr
\noalign{\smallskip}
\noalign{\hrule}
\noalign{\smallskip}
BATSE Hardness Ratio $> 10$ (0.05--0.3 MeV)&21&1\cr
COMPTEL Detections (1--30 MeV)&4&20\cr
OSSE Detections (0.06--10 MeV)&0&2\cr
EGRET Detections (20-30,000 MeV)&0&6\cr}
\bigskip
Table 1: BATSE hardness and high energy detections; nominal sensitivity
ranges are indicated.  Data are from 3B Catalogue$^3$.
\par
\vfil
\eject
\bye
\end